\documentclass{article}
\usepackage{ijcai17}

\usepackage{times}
\usepackage{latexsym}
\usepackage{comment}
\usepackage{graphicx}
\usepackage{graphics}
\usepackage{epsfig}
\usepackage{subfigure}
\usepackage[cmex10]{amsmath}
\usepackage{subfigure}
\usepackage{floatrow}
\usepackage{diagbox}
\newfloatcommand{capbtabbox}{table}[][\FBwidth]

\title{Learning Spatiotemporal-Aware Representation for POI Recommendation}
\author{Bei Liu{\small $~^{1}$}, Tieyun Qian{\small $~^{1}$}, Bing Liu{\small $~^{2}$}, Liang Hong{\small $~^{3}$}, Zhenni You{\small $~^{1}$}, Yuxiang Li{\small $~^{1}$}\\
{\small $~^{1}$} State Key Laboratory of Software Engineering, Wuhan University, China\\
{\small $~^{2}$} Department of Computer Science, University of Illinois at Chicago, USA  \\
{\small $~^{3}$} School of Information Management,  Wuhan University, China\\
\{qty, beiliu\}@whu.edu.cn, liub@cs.uic.edu, \{hong, znyou, liyux\}@whu.edu.cn\\
}
\begin{document}

\maketitle

\begin{abstract}
The wide spread of location-based social networks brings about a huge volume of user check-in data, which facilitates the recommendation of points of interest (POIs). Recent advances on distributed representation shed light on learning low dimensional dense vectors to alleviate the data sparsity problem. Current studies on representation learning for POI recommendation embed both users and POIs in a common latent space, and users' preference is inferred based on the distance/similarity between a user and a POI. Such an approach is not in accordance with the semantics of users and POIs as they are inherently different objects. In this paper, we present a novel spatiotemporal aware (STA) representation, which models the spatial and temporal information as \emph{a relationship connecting users and POIs}. Our model generalizes the recent advances in knowledge graph embedding. The basic idea is that the embedding of a $<$time, location$>$ pair corresponds to a translation from embeddings of users to POIs. Since the POI embedding should be close to the user embedding plus the relationship vector, the recommendation can be performed by selecting the top-\emph{k} POIs similar to the translated POI, which are all of the same type of objects.
We conduct extensive experiments on two real-world datasets. The results demonstrate that our STA model achieves the state-of-the-art performance in terms of high recommendation accuracy, robustness to data sparsity and effectiveness in handling cold start problem.
\end{abstract}

\section{Introduction}
Location-based social networks (LBSN), such as Foursquare, Yelp, and Facebook Places, are becoming pervasive in our daily lives. Users on LBSN like to share their experiences with their friends for points of interest (POIs), e.g., restaurants and museums. The providers of location-based services have collected  a huge amount of users' check-in data, which facilitates the recommendation of POIs to unvisited users. The POI recommendation is of high value to both the users and companies, and thus has attracted much attention from researchers in recent years \cite{Cheng_tist16,Zhu_kdd15,Chen_aaai15,Gao_aaai15}.

Most existing studies mainly focused on leveraging spatial information due to the well-known strong correlation between users' activities and geographical distance ~\cite{Zheng_www09,Ye_gis10,Cho_kdd11}. For example, Ye et al.~\shortcite{Ye_sigir11}  proposed a Bayesian collaborative filtering (CF) algorithm to explore the geographical influence. Cheng et al. ~\shortcite{Cheng_aaai12} captured the geographical influence by modeling the probability of a user's check-in on a location as a multi-center Gaussian model and then combined it into a generalized matrix factorization model.  Lian et al. ~\shortcite{Lian_kdd14} adopted a weighted matrix factorization framework to incorporate the spatial clustering phenomenon.

Similar to the geo-spatial information, time is another important factor in POI recommendation. Ye et al., ~\shortcite{Ye2011_gis11} found the periodic temporal property that people usually go to restaurants at around noon and visit clubs at night.  Yuan et al., ~\shortcite{Yuan_sigir13} developed a CF based model to integrate temporal cyclic patterns. Cheng et al. ~\cite{Cheng_ijcai13} explored the temporal sequential patterns for personalized POI recommendation by using the transition probability of two successive check-ins of a user.

Existing studies has exploited spatial or temporal influences mainly using CF ~\cite{Ye_sigir11,Yuan_sigir13} and Markov transition approaches~\cite{Cheng_ijcai13}. Due to the sparsity of users' check-in records, it is hard to find similar users or calculates transition probability.  Although matrix factorization (MF) methods are effective in dealing with the sparsity in user-POI matrix ~\cite{Cheng_aaai12,Lian_kdd14}, they do not consider the current location of the user. More importantly, while time and location together play a critical role in determining users' activities in LBSNs, rare work has modeled their joint effects. Considering only one factor will deteriorate the predictive accuracy. For instance, a student may go to a school cafeteria or to a food court in a mall at lunch time depending on he/she is on campus or outside. It is not suggested for a system to recommend the same restaurant to a user at the same time but different location. This example shows the ineffectiveness when using one type of information but ignoring the other. However, taking both time and location into consideration exaggerates the data sparsity.

In this paper, we propose a novel spatiotemporal aware (STA) model, which captures the joint effects of spatial and temporal information. Our model has the following distinct characteristics.
\begin{itemize}
  \item  STA takes location and time as a whole to determine the users' choice of POIs.
  \item  STA embeds a spatiotemporal pair $<$time, location$>$ as \emph{a relationship} connecting users and POIs.
\end{itemize}
By considering the time and location at the same time, our model can be successfully applied to real-time POI recommendation. Furthermore, distributed representations of STA are very effective in solving the problem of data sparsity.

Two recent works ~\cite{Feng_ijcai15,Xie_cikm16} also exploited the power of distributed representation for alleviating data sparsity. The personalized ranking metric embedding  (PRME) by Feng et al. ~\shortcite{Feng_ijcai15} projected each POI and each user into a latent space, and then recommended a POI $v$ to a user $u$ at location $l$ based on the Euclidean distance between the POI and the user $\parallel\vec{u}-\vec{v}\parallel^2$ and that between the POI and the location $\parallel\vec{l}-\vec{v}\parallel^2$. Xie et al.~\shortcite{Xie_cikm16} proposed a graph based embedding model (GE) by embedding graphs into a shared low dimensional space, and then computed the similarity between a user $u$'s query $q$ at current time $t$ and location $l$  and a POI $v$ using an inner product, $S(q,v)=\vec{u}^T\cdot\vec{v}+\vec{t}^T\cdot\vec{v}+\vec{l}^T\cdot\vec{v}$. While PRME, especially GE, shows significant improvements over many other baselines, these two methods have the drawback that they embed both users and POIs in a common latent space, and users' preference is inferred based on the distance/similarity between a user and a POI. Such an approach is unnatural since users and POIs are inherently different objects. In contrast, our STA model generalizes recent advances in knowledge graph embedding ~\cite{Lin_aaai15}. A user $u$ reaches an interested POI $v_q$ via an edge $tl$ denoting the $<$time, location$>$ pair, i.e., $\vec{u} + \vec{tl} \approx \vec{v_q}$. With this transformation, we can do recommendation for $u$ by selecting the top-\emph{k} POIs similar to POI $v_q$, which are all of the same type of objects with similar semantics.

\section{Problem Definition and Preliminary}
\quad Definition 1. (\emph{\textbf{POI}}) A POI \emph{v} is defined as a unique identifier representing one specific position (e.g., a cafe or a hotel), and \emph{V} is a set of POIs, i.e., $V=\{v|v=(pid, position)\}$. 

\quad Definition 2. (\emph{\textbf{Check-in Activity}}) A check-in activity is a quadruple (u, t, l, v), which means a user \emph{u} visits a POI \emph{v} in location \emph{l} at time \emph{t}.

Definition 3. (\emph{\textbf{Spatiotemporal pattern}}) A spatiotemporal pattern, denoted as \emph{tl}, is a combination of a time slot \emph{t} and a location \emph{l} like $<$11 a.m., Los Angeles$>$.
%For example, we can get a pattern $<$11 a.m., Chicago$>$, if there is a check-in whose time is 11 o'clock in the morning and location is Chicago. We use \emph{tl} to indicate a TL identifier.

%Definition 4. (\emph{\textbf{TL-translation}}) We define a TL-translation as the connection between user \emph{u} and POI \emph{v} corresponding to a spatiotemporal pattern. More specifically,   a TL-translation means in this situation (time \emph{t} and location \emph{l}) \emph{u} tends to visit\emph{ v}.

For ease of presentation, we summarize the notations in Table \ref{tbl:note}.  The POI recommendation problem investigated in this paper has the same settings as that in ~\cite{Xie_cikm16}.  The formal problem definition is as follows.
\begin{table}
\small
\begin{tabular} {c|c}
\hline
Variable	& Interpretation\\
\hline
\emph{u}, \emph{v} & the user \emph{u} and POI \emph{v} \\
\hline
\emph{t} & the time slot discretized from timestamp\\
\hline
\emph{l} & the location mapped from (longitude, latitude)\\
\hline
\emph{tl} & the spatiotemporal pattern $<$t, l$>$ \\
\hline
$\vec{u}$,$\vec{tl}$,$\vec{v}$ & embeddings of \emph{u}, (\emph{t,l}), and \emph{v}\\
\hline
$u_q$, $t_q$, $l_q$ & query user $u_q$, his/her current time $t_q$ and location $l_q$\\
\hline
$v_q$  & the potential POI that query user $u_q$ is interested in\\
\hline
%\emph{D} & users' activity set $D=\{d|d=(u,t,l,v)\}$\\
%\hline
%\emph{U}, \emph{V} & the user set \emph{U} and POI set \emph{V} \\
%\hline
\end{tabular}
\caption{Notations used in this paper}
\label{tbl:note}
\vspace{-0.2cm}
\end{table}

\textbf{Problem Definition} (Location-based Recommendation) Given a dataset $D=\{d|d=(u,t,l,v)\}$ recording a set of users' activities, and a query $q=(u_q, t_q, l_q)$, we aim to recommend top-\emph{k} POIs in \emph{V} that the query user $u_q$ would be interested in.

\textbf{Preliminary - KG Embedding}  The knowledge graph (KG) is a directed graph whose nodes and edges describing \emph{entities} and their \emph{relations} of the form (\emph{head, relation, tail}), denoted as (\emph{h, r, t}). The goal of knowledge graph embedding is to learn a continuous vector space where the embeddings of entity and relation can preserve certain information of the graph.
Bordes et al.~\shortcite{Bordes_ml14} presented a simple yet effective approach TransE to learn vector embeddings for both entities and relations in KG. The basic idea is that the relationship between entities corresponds to a translation the embeddings of entities, namely, $\vec{h}+\vec{r}\approx\vec{t}$ when (\emph{h ,r, t}) exits in graph. Later, a model named TransH~\cite{Wang_aaai14} was proposed to enable an entity to have distinct representations when it is involved in different relations.

Both TransE and TransH project all entities and relations into the same space. However, some entities may have multiple aspects and relations focusing on different aspects of the entities. Such entities are close in the entity space when they are similar, but they should be far away from each other in the relation space if they are strongly different in some specific aspects. To address this issue, Lin et al.~\shortcite {Lin_aaai15} presented a TransR model to project two entities \emph{h} and \emph{r} of (\emph{h,r,t}) into a \emph{r}-relation space as $h_r$ and $t_r$ with operation $M_r$, such that $\vec{h_r}+\vec{r}\approx\vec{t_r}$ holds in the relation space.

\section{Our Proposed Framework}
We seek to learn the representations with the following characteristics.
\begin{itemize}
  \item Spatiotemporal awareness - Location and time together play a crucial role when a user selects a POI; they should not be separated into individual ones.
  \item Semantics consistency - All the POIs, either the query user's interested POI $v_q$ or all existing POIs $v \in V$, should come from a consistent semantic space.
\end{itemize}

In order to satisfy the first requirement, we combine each time slot and location as a spatiotemporal pattern $<$\emph{t}, \emph{l}$>$, and convert the  quadruples $(u,t,l,v) \in D$ into triples ($u$, $<$$t$, $l$$>$, $v$) in $D'$. We then learn representations for users, spatiotemporal patterns, and POIs from the converted set $D'$ to meet the second condition, using the translation technique originated from knowledge graph embedding.

\subsection{STA model}
\begin{figure}[htb]
\vspace{-0.3cm}
\centerline{\includegraphics[width=0.68\textwidth,height=4.2cm]{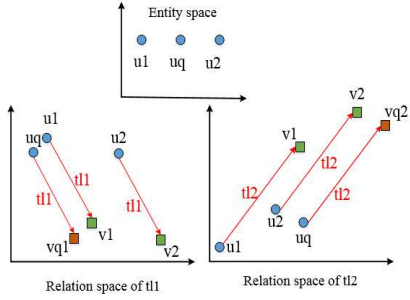}}
\vspace{-0.2cm}
\caption{Impacts of spatiotemporal patterns}
\label{fig:relation}
%\vspace{-0.2cm}
\end{figure}

For the location-based recommendation problem, we focus on the connections between users and POIs corresponding to the spatiotemporal relations. Intuitively, if a POI \emph{v} is often visited by similar users in location \emph{l} at time \emph{t}, the probability of a  query user $u_q$ visiting \emph{v} with the same spatiotemporal relation will be high. On the other hand, users similar in the entity space may visit different POIs under distinct temporal and geographic conditions.
%On the other hand, if the user $u_q$ at location \emph{l} visited POI \emph{v} very often at time \emph{t}, he/she tends to visit \emph{v} again under the same spatiotemporal condition.
In order to capture the strong correlations of users and POIs to the spatiotemporal patterns, we generalize the TransR technique ~\cite{Lin_aaai15} to fit the POI recommendation task.
The basic idea is that a user $u$ will reach an interested POI $v_q$ via a translation edge $tl$, i.e., $\vec{u} + \vec{tl} \approx \vec{v_q}$.
Fig.~\ref{fig:relation} illustrates the impacts of \emph{tl} patterns.

In Fig.~\ref{fig:relation}, suppose $u_1$, $u_q$, and $u_2$ are three university students,  $u_1$ and $u_q$ taking same courses, and $u_2$ and $u_q$ sharing the dormitory. Given two patterns $tl_1=<12 a.m., campus>$ and $tl_2=<8 p.m., dormitory>$, the query user  $u_q$ will be translated into two POIs $v_{q1}$ and $v_{q2}$, hence we should recommend for $u_q$ the POI $v_1$ in the left lower sub-figure and $v_2$ in the right lower sub-figure, which are the close neighbor of  $v_{q1}$ and $v_{q2}$, respectively. The different recommending results $v_1$ and $v_2$ are caused by the effects of different spatiotemporal relations $tl_1$ and $tl_2$.

We now introduce the detail for STA. For each triple ($u$, $<$$t$, $l$$>$, $v$) in $D'$, the user $u$, the spatiotemporal pair $<$$t$, $l$$>$ ($tl$ in short), and POI $v$ corresponds to the head entity \emph{h}, the relationship edge \emph{r} and the tail entity \emph{t} in TransR, respectively. Their embeddings are set as $\vec{u}$, $\vec{v}$ $\in$ $\Re^d$, and $\vec{tl}$ $\in$ $\Re^m$. For each spatiotemporal pair $tl$, we set a projection matrix $M_{tl} \in \Re^{d\times m}$ to project a user embedding $\vec{u}$ and a POI embedding $\vec{v}$ in the original entity space to $\vec{u_{tl}}=\vec{u}M_{tl}$ and $\vec{v_{tl}}=\vec{v}M_{tl}$ in the relation space, such that $\vec{u_{tl}} + \vec{tl} \approx \vec{v_{tl}}$. This indicates that a POI embedding $\vec{v_{tl}}$ should be the nearest neighbor of $\vec{u_{tl}} + \vec{tl}$. Hence the the score function can be defined as:
\begin{eqnarray}\label{score}
\small
  \nonumber s_{tl}(u, v) = \parallel \vec{u_{tl}} + \vec{tl} - \vec{v_{tl}} \parallel _2^2\\
  \nonumber   s.t. \parallel \vec{u} \parallel _2 \leq 1, \parallel \vec{v} \parallel _2 \leq 1, \parallel \vec{tl} \parallel _2 \leq 1,  \\
   \parallel \vec{u_{tl}} \parallel _2 \leq 1, \parallel \vec{v_{tl}} \parallel _2 \leq 1
\end{eqnarray}

Given the score function defined in Eq. \ref{score} for a triple ($u$, $tl$, $v$), the entire objective function for training is as follows.
\begin{equation}\label{obj}
\small
    L = \sum_{(u,tl,v) \in T} \sum_{(u',tl,v') \in T'} max(0, s_{tl}(u, v) + \gamma - s_{tl}(u', v')),
\end{equation}
where max(\emph{a},\emph{b}) is used to get the maximum between \emph{a} and \emph{b}, $\gamma$ is the margin, \emph{T} and \emph{T'} are the sets of correct and corrupted triples, respectively. The corrupted triples are generated by replacing the head and tail entities in correct triples using the same sampling method as that in~\cite{Wang_aaai14}.

We adopt stochastic gradient descent (SGD) (in mini-batch mode) to minimize the objective function in Eq. \ref{obj}. A small set of triplets, is sampled from the training data.  For each such triplet, we sample its corresponding incorrect triplets. All the correct and incorrect triples are put into a mini-batch. We compute the gradient and update the parameters after each mini-batch. When the iteration reaches a predefined number, we learn all the embedding for users, POIs, and spatiotemporal patterns.

\subsection{Recommendation Using STA}
Once we have learned the embeddings, given a query user $u_q$ with the query time $t_q$ and location $l_q$, i.e., \emph{q} = ($u_q$, $t_q$, $l_q$), we first combine $t_q$ and $l_q$ as a spatiotemporal pattern $tl_q$, and then we can get the potential POI $v_q$ using Eq. \ref{querypoi}.
\begin{equation}\label{querypoi}
\small
   %\vec{v_{tlq}} = \vec{u_{tlq}} + \vec{tl_q}
   \vec{v_q} = \vec{u_q} M_{tl} + \vec{tl_q}
\end{equation}

The learned POI embedding $v_q$ naturally reflects the user's preference, because it encodes the users' past activities in $\vec{u_q}$. It also captures the geographic and temporal influence in $\vec{tl_q}$.

For each POI $v \in V$, we compute its distance to the POI $v_q$ in the normed linear space as defined in Eq. ~\ref{simpoi}, and then select the \emph{k} POIs with the smallest ranking scores as recommendations.
\begin{equation}\label{simpoi}
\small
   %d(v_{tl}, v_{tlq}) = \parallel \vec{v_{tl}} - \vec{v_{tlq}} \parallel_1
   d(v, v_q) = \parallel \vec{v} M_{tl} - \vec{v_q} \parallel_1
\end{equation}

We would like to emphasize our differences in computing $v_q$ and recommending POIs from those in ~\cite{Lin_aaai15,Xie_cikm16}. First, we can find an explicit POI $v_q$ directly from the latent space through the translation of the embedding of the spatiotemporal pattern on the user's embedding, while others compute an implicit $v_q$ by its distance/similarity to user $u_q$. Second, since the embeddings for POIs in \emph{V} are also from the same space, we can choose the ones which are the closest neighbors of $v_q$ in this space. This indicates that our recommended POIs are semantically consistent with the query user's interested POI $v_q$.

\subsection{Dealing with Cold Start POIs}
%The cold start POIs refer to those containing tag information but without any check-ins ~\cite{Xie_cikm16}.
Considering the cold start POIs, which contain geographic and content information like tags but do not have any check-ins~\cite{Xie_cikm16}, we can simply extend our model to include the POI-POI relationship through the translation of content patterns. We call this model STA-C. The rationale is that, if two POIs share a common tag or location, there will be a high degree of similarity between them, and their vector representations should be close to each other. Based on this observation, we define the score function as following:
\begin{comment}
\begin{eqnarray}\label{score}
  \nonumber s_{tlw}(u, v, vs) = s_{tl}(u, v) + s_{w}(v, vs) \\
    = \parallel \vec{u_{tl}} + \vec{tl} - \vec{v_{tl}} \parallel _2^2 + \parallel \vec{v_{w}} + \vec{w} - \vec{vs_{w}} \parallel _2^2,
  \end{eqnarray}
\end{comment}

\begin{equation}\label{scorecold}
\small
\begin{aligned}
\begin{split}
s_{tlw}(u, v, s)  & = s_{tl}(u, v) + s_{wl}(v, s) \\
    %& = s_{tl}(u, v) + s_{w}(v, s) \\
    & = \parallel \vec{u_{tl}} + \vec{tl} - \vec{v_{tl}} \parallel _2^2 + \parallel \vec{v_{wl}} + \vec{wl} - \vec{s_{wl}} \parallel _2^2,
\end{split}
\end{aligned}
\end{equation}

where \emph{s} is a POI sharing at least one $<$word, location$>$ pair with POI \emph{v}, and the objective function for cold start POIs is defined as:
\begin{comment}
\begin{eqnarray}\label{objcold}
    \nonumber L = \sum_{(u,tl,v) \in T} \sum_{(u',tl,v') \in T'} max(0, s_{tl}(u, v) + \gamma - s_{tl}(u', v')) \\
    + \sum_{(v,w,s) \in W} \sum_{(v',w,s') \in W'} max(0, s_{w}(v, s) + \gamma - s_{w}(v', s'))
\end{eqnarray}
\end{comment}

\begin{equation}\label{objcold}
\small
\begin{aligned}
\begin{split}
    LC = & \sum_{(u,tl,v) \in T} \sum_{(u',tl,v') \in T'} max(0, s_{tl}(u, v) + \gamma - s_{tl}(u', v')) + \\
    & \sum_{(v,wl,s) \in W} \sum_{(v',wl,s') \in W'} max(0, s_{wl}(v, s) + \gamma - s_{wl}(v', s'))
\end{split}
\end{aligned}
\end{equation}

We once again use stochastic gradient descent to minimize the objective function \emph{LC} in Eq. \ref{objcold}. The only difference is the sampling procedure. For STA-C, since we have two types of edges, we  sample the triplets (u, tl, v) and (v, wl, s) and their corresponding incorrect triples alternatively to update the model.

Our STA-C model proposed for dealing with cold start POIs can also be applied to the normal POI recommendation problem. However, it requires that those POIs should contain content information. For the recommendation on datasets like Gowalla, STA-C is not valid. Hence we only treat it as an extended model. Please also note that, it is STA-C that uses the same information as GE does. Our standard STA model, on the other hand, uses less information than GE because it does not include the contents of POIs.

\section{Experimental Evaluation}
In this section, we first introduce the experimental setup and then compare our experimental results with those of baselines. Finally we show the performance of our method for addressing the data sparsity and cold start problem.
\subsection{Experimental Setup}
\textbf{Datasets}
We evaluate our methods on two real-life LBSN datasets: Foursquare and Gowalla. A number of researchers have conducted experiments on data collected from these two social networks ~\cite{Yuan_sigir13,Chen_aaai15,Gao_aaai15,Xie_cikm16,Yin_tkde16}. However, many of them are collected from various regions or in different time spans. For a fair comparison with GE, we use the publicly available version \footnote{https:/sites.google.com/site/dbhongzhi} provided by the authors of ~\cite{Xie_cikm16}.

The two datasets have different scales such as geographic ranges, the number of users, POIs, and check-ins.  Hence they are good for examining the performance of algorithms on various data types. Their statistics are listed in Table~\ref{tbl:dataset}.
\begin{table}[!htb]
\small
\begin{tabular} {ccc}
%\hline
& \textbf{Foursquare} & \textbf{Gowalla} \\
\hline
\# of users & 114,508 & 107,092 \\
%\hline
\# of POIs  & 62,462 & 1,280,969 \\
%\hline
\# of Check-ins  & 1,434,668 & 6,442,892 \\
%\hline
\#std time slots  & 24   & 24   \\
%\hline
\# of locations & 5,846 & 200\\
%\hline
\# of $<$t, l$>$ patterns &  28,868 & 3,636\\
\hline
\end{tabular}
\caption{Statistics of two datasets} \label{tbl:dataset}
\end{table}

Each check-in is stored as user-ID, POI-ID, POI-location in the form of latitude and longitude, check-in timestamp, and POI-content (only for Foursquare).
In order to get the spatiotemporal patterns $<$t, l$>$ in Table~\ref{tbl:dataset}, we use the same discretized method as that in ~\cite{Xie_cikm16}, i.e., dividing time into 24 time slots which correspond to 24 hours, and the whole geographical space into a set of regions according to 5,846 administrative divisions (for Foursquare) and 200 regions clustered by a standard \emph{k}-means method (for Gowalla). We finally get 28,868 and 3,636 $<$t, l$>$ pairs on Foursquare and Gowalla, respectively.

\textbf{Baselines - \{GE, STA-E, STA-H\}}
We use GE, the state-of-the-art location based recommendation approach in ~\cite{Xie_cikm16}, as our baseline. GE adopts a graph-based embedding framework. It learns the embeddings based on the POI-POI, POI-Time, POI-Location, and POI-Words graphs. By integrating the sequential, geographical, temporal cyclic, and semantic effect into a shared space, GE effectively overcomes the data sparsity problem and reaches the best performance so far.

We do not compare our method with other existing approaches because, GE has already significantly outperformed a number of baselines including JIM ~\cite{Yin_cikm15joint}, PRME ~\cite{Feng_ijcai15}, and Geo-SAGE ~\cite{Wang_kdd15Geo}. We thus only show our improvements over GE.

Also note that although we choose the TransR technique in knowledge graph embedding to materialize our STA model, the essential of our proposed framework is the translation of $<$time, location$>$ pairs in the embedding space. This indicates that we do not rely on a specific translation model. Hence we can use TransE ~\cite{Bordes_ml14} and TransH ~\cite{Wang_aaai14} to realize STA. We denote the resulting methods as STA-E and STA-H baselines, respectively.

\textbf{Settings}
We first organize the quadruples (u, v, t, l) in each dataset by users to get each user's profile $D_u$. We then rank the records in $D_u$ according to the check-in timestamps, and finally divide these ordered records into two parts: the first 80\% as the training data, and the rest 20\% data as the test data. Moreover, the last 10\% check-in records in the training data are used as a validation set for tuning the hyper-parameters. We use the accuracy@\emph{k} (\emph{k} = \{1, 5, 10, 15, 20\}) as our evaluation metric. All these settings, as well as the computation approach to accuracy@\emph{k}, are same as those in ~\cite{Xie_cikm16}.

We use the default settings in the original TransR ~\cite{Lin_aaai15} as the parameter settings for our STA model. Specifically, we set the learning rate $\lambda = 0.0001$, the margin $\gamma = 2$, the mini-batch size $B=4800$, and the embedding dimensions $m = d = 100$, and we traverse over all the training data for 1000 rounds.

\subsection{Comparison with baselines}
\begin{figure*}[htb]
\centerline{\includegraphics[width=0.82\textwidth,height=5.2cm]{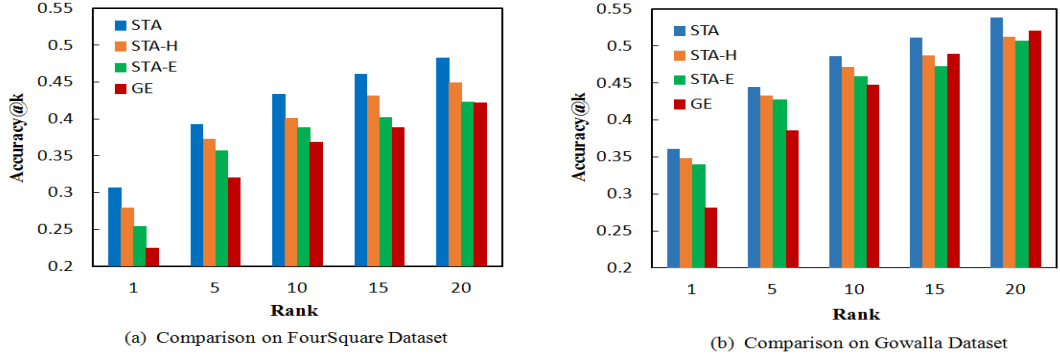}}
\caption{Comparisons with baselines}
\label{fig:comp}
\vspace{-0.1cm}
\end{figure*}

For a fair comparison, we implement GE using the same LINE software provided by the authors of ~\cite{Tang_www15} on our data divisions. All the parameters for GE are same as those in ~\cite{Xie_cikm16}. We find a slightly difference (less than 1\% in accuracy) between the original results and those by our implemented GE. This is understandable and acceptable considering the randomness when sampling negative edges in LINE and initiating the centers of clusters of regions. All parameters for STA-E and STA-H use the default settings in ~\cite{Bordes_ml14} and ~\cite{Wang_aaai14}.
We present the comparison results on Foursquare and Gowalla in Fig.~\ref{fig:comp} (a) and (b), respectively.

\begin{comment}
\begin{figure}[htb]
\centerline{\includegraphics[width=0.46\textwidth,height=3.5cm]{compf.eps}}
\caption{Comparisons on Foursquare}
\label{fig:compf}
\end{figure}
\begin{figure}[htb]
\centerline{\includegraphics[width=0.46\textwidth,height=3.5cm]{compw.eps}}
\caption{Comparisons on Gowalla}
\label{fig:compw}
\end{figure}
\end{comment}

From Fig.~\ref{fig:comp} (a), it is clear that all our proposed STA-style models significantly outperform GE. For instance, the accuracy@1 for STA, STA-H, and STA-E is 0.307, 0.280, 0.255, respectively, much better than 0.225 for GE. Similar results can be observed in Fig. ~\ref{fig:comp} (b) on Gowalla dataset. This clearly demonstrates the effectiveness of our translation based framework.

While STA shows drastic improvement over GE for all \emph{k}s on Foursquare, the trend is not that obvious on Gowalla when \emph{k} = 15, 20.
This is because there is a much smaller number of relations in Gowalla than that in Foursquare. As shown in Table ~\ref{tbl:dataset}, Gowalla only has 3,636 relation patterns ($<$t, l$>$ pairs) while Foursquare has 28,868 pairs. Hence the learnt embeddings for entities and relations are worse than those on Foursquare, and incur the less accurate results when \emph{k} is large.

Besides the improvement over GE, STA outperforms STA-H and STA-E as well. The reason is that TransR can differentiate the entities in the transformed relation space. Nevertheless, we see a less significant enhancement of STA over STA-H on Gowalla. This also conforms to the characteristics of the data: the graph of Gowalla is much larger but has less \emph{tl} relation edges than that of Foursquare, and the advantage of TransR over TransE is not obvious on such a dataset.

\subsection{Effects of Model Parameters}
%There are three main parameters involved in GE, i.e., the  embedding dimension \emph{d}, the number of samples \emph{N} in LINE ~\cite{Tang_www15}, and the time interval.
The effects of embedding dimension \emph{d} on Foursquare and Gowalla are shown in Table~\ref{tab:dimf} and Table~\ref{tab:dimw}, respectively.
\begin{table}[!htb]
{
\small
\caption{Effects of Dimensionality on Foursquare}
       \begin{tabular}{|c|c|c|c|c|c|}
       \hline
        \diagbox{\emph{d}}{Acc}{\emph{k}} & 1 & 5 & 10 & 15 & 20\\
        \hline
        70 & 0.281 & 0.376 & 0.409 & 0.433 & 0.451\\
        \hline
        80 & 0.294 & 0.384 & 0.417 & 0.445 & 0.462\\
        \hline
        90 & 0.300 & 0.390 & 0.425 & 0.459 & 0.476\\
        \hline
        100 & 0.307 & 0.393 & 0.434 & 0.461 & 0.483\\
        \hline
        110 & 0.311 & 0.407 & 0.439 & 0.463 & 0.486\\
        \hline
        120 & 0.312 & 0.407 & 0.439 & 0.464 & 0.486\\
        \hline
       \end{tabular}
       \label{tab:dimf}
}
\end{table}

\begin{table}
{
\small
\caption{Effects of Dimensionality on Gowalla}
       \begin{tabular}{|c|c|c|c|c|c|}
       \hline
        \diagbox{\emph{d}}{Acc}{\emph{k}} & 1 & 5 & 10 & 15 & 20\\
        \hline
        70 &  0.355 & 0.432 & 0.474 & 0.503 & 0.527\\
        \hline
        80 & 0.358 & 0.436 & 0.478 & 0.508 & 0.530\\
        \hline
        90 & 0.359 & 0.439 & 0.482 & 0.509 & 0.535\\
        \hline
        100 & 0.361 & 0.445 & 0.486 & 0.511 & 0.539\\
        \hline
        110 & 0.361 & 0.445 & 0.488 & 0.513 & 0.540\\
        \hline
        120 & 0.361 & 0.445 & 0.488 & 0.513 & 0.540\\
        \hline
       \end{tabular}
       \label{tab:dimw}
}
\end{table}

We can see that the experimental results are not very sensitive to the dimension \emph{d}. With an increasing number of dimension, the accuracy on Gowalla is almost unchanged, i.e., the improvement is less than 1\% in nearly all cases. The accuracy on Foursquare is slightly enhanced with a large dimension \emph{d}, and finally it becomes stable.

\begin{figure*}[!htb]
\centerline{\includegraphics[width=0.82\textwidth,height=5.5cm]{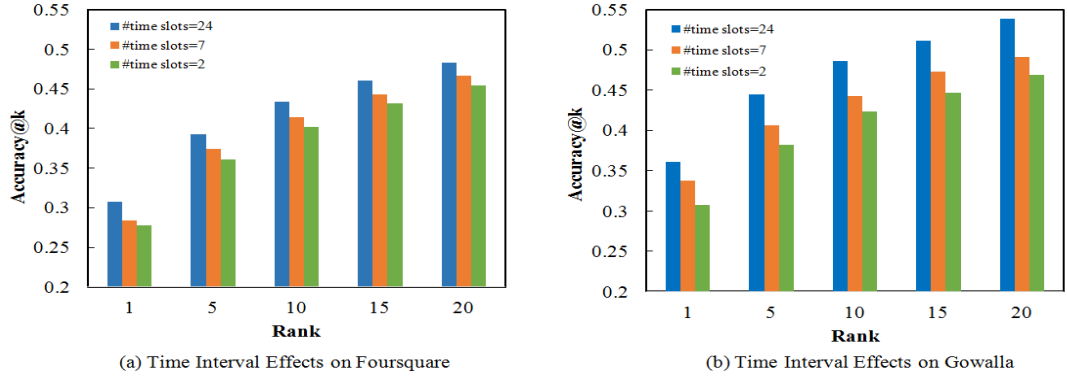}}
\caption{Effects of Time Interval}
\label{fig:time}
\end{figure*}

To investigate the effects of time interval, we divide timestamps by three methods, i.e., splitting time into 24, 7, and 2 time slots, corresponding to the daily, weekly, and weekday/weekend patterns, respectively. Figure~\ref{fig:time} shows the effects of various time intervals.
\vspace{-0.0cm}
We observe that the impact of the daily patterns is the most significant on both datasets. In addition, the results for different patterns vary widely, suggesting a good strategy for dividing the time slot is important.

\subsection{Sensitivity to Data Sparsity}
To investigate the sensitivity to data sparsity of STA and GE, we conduct extensive experiments to evaluate the performance on two datasets by reducing training data. More precisely, we keep the testing dataset unchanged and reduce the training data randomly by a ratio of 5\% to 20\% stepped by 5. Due to the space limitation, we only present the results by reducing 20\% training data Table~\ref{tab:20less}. The trends with other ratios are all alike.

\vspace{-0.0cm}
\begin{table*}[!htb]
\small
\centering
\subtable[ on Foursquare]{
       \begin{tabular}{|c|c|c|c|c|c|c|}
       \hline
        &\multicolumn{3}{c|}{GE} & \multicolumn{3}{c|}{STA}  \\
        \hline
        \emph{k} &GE    & GE- & change & STA   & STA- & change\\
        \hline
        1 &0.225 & 0.154 & -31.69\% & 0.307 & 0.246 & -20.00\% \\
        \hline
        5 &0.321 & 0.228 & -28.84\% & 0.393 & 0.320 & -18.46\% \\
        \hline
        10 &0.369 & 0.270 & -26.82\% & 0.434 & 0.365 & -15.86\% \\
        \hline
        15 &0.388 & 0.295 & -23.95\% & 0.461 & 0.382 & -17.04\% \\
        \hline
        20 &0.422 & 0.318 & -24.68\% & 0.483 & 0.407 & -15.73\% \\
        \hline
        \end{tabular}
       \label{tab:f20}
}
\quad
\subtable[ on Gowalla]{
       \begin{tabular}{|c|c|c|c|c|c|c|}
       \hline
        &\multicolumn{3}{c|}{GE} & \multicolumn{3}{c|}{STA}  \\
        \hline
        \emph{k} &GE    & GE- & change & STA   & STA- & change\\
        \hline
        1 & 0.282 & 0.209 & -25.77\% & 0.361 & 0.291 & -19.36\%\\
        \hline
        5 & 0.386 & 0.303 & -21.67\% & 0.445 & 0.384 & -13.72\%\\
        \hline
        10 & 0.448 & 0.354 & -20.98\% & 0.486 & 0.415 & -14.64\%\\
        \hline
        15 & 0.489 & 0.396 & -19.12\% & 0.511 & 0.445 & -13.01\%\\
        \hline
        20 & 0.521 & 0.423 & -18.91\% & 0.539 & 0.468 & -13.14\%\\
 \hline
       \end{tabular}
       \label{tab:w20}
}
\caption{Sensitivity to Sparsity (GE- and STA- for 20\% less training data)}
\label{tab:20less}
\end{table*}

We have the following important notes for Table~\ref{tab:20less}.
%\begin{comment}
\begin{itemize}
  \item  With the reduction of training data, the accuracy values for STA and GE both decrease. However, STA always achieves the best results at different \emph{k} values on two datasets.
    \item The reduction of accuracy of our STA model  is much smaller than that of GE. For instance, the accuracy@1 of GE decreases from 0.225 to 0.154, showing a 31.69\% drop. In contrast, our STA model only has a 20.00\% change. This strongly suggests that our model is more robust to the data sparsity.
    \item The declination of accuracy on Foursquare is more obvious than on Gowalla. The reason may be that Foursquare is much sparser in users' check-ins than Gowalla, hence reducing the training data has a greater impact on Foursquare.
\end{itemize}
%\end{comment}

\subsection{Test for Cold Start Problem}
\begin{figure}
\centerline{\includegraphics[width=0.56\textwidth,height=3.6cm]{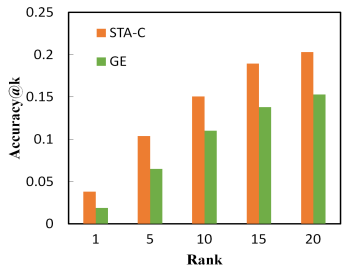}}
\vspace{-0.1cm}
\caption{Test for Cold Start Problem on Foursquare}
\label{fig:cold}
\vspace{-0.0cm}
\end{figure}

In this experiment, we further compare the effectiveness of our extended  STA-C model with GE when addressing the cold-start problem. The cold start POIs are defined as those visited by less than 5 users ~\cite{Yin_tkde16}. To test the performance of cold start POI recommendations, we select users who have at least one cold-start check-in as test users. For each test user, we choose his/her check-in records associated with cold-start POIs as test data and the remains as training data.
Since there is no content information for POIs in Gowalla, we conduct experiments, just as GE did, only on Foursquare. The results are shown in Fig.~\ref{fig:cold}.
%\vspace{-0.2cm}

From Fig.~\ref{fig:cold}, it is clear that our proposed STA-C model consistently beats GE when recommending cold start POIs. The superior performance of STA-C model is due to the translation of content and geography information \emph{wl} from an ordinary POI \emph{v} to a cold start POI $v_c$. As long as there is an existing \emph{v} sharing one $<$word, location$>$ pair with $v_c$, our STA-C model can get a translation for $v_c$. In contrast, GE utilizes the bipartite graphs of POI-Word and POI-Location. The weight of an edge in the graph is calculated by a TF-IDF value of the word or the frequency of a location. The edge weight is proportional to the probability of edge sampling. Since there are few check-in records for cold start POIs, a $v_c$-word and $v_c$-location edge has an extremely rare chance to be selected and updated. Consequently, the learnt embedding for $v_c$ will be poor and further deteriorates the recommendation accuracy.

\section{Conclusion}
We present a novel spatiotemporal aware model STA for learning representations of users, spatiotemporal patterns, and POIs. The basic idea is to capture the  geographic and temporal effects using a $<$time, location$>$ pair, and then model it as a translation connecting users and POIs. We realize STA using the knowledge graph embedding technique. Our method has two distinguished advantages. 1) We learn a joint representation for spatiotemporal patterns whose components contribute together to a user's choice in POIs. 2) The translation mechanism enables the learnt POI embeddings to be in the same semantic space with that of the query POI.

We conduct extensive experiments on two real-life datasets. Our results show that STA achieves the state-of-the-art performance in recommendation accuracy. It also significantly outperforms the baselines in terms of the effectiveness in addressing both the data sparsity and cold start problems.

\section*{Acknowledge}
The work described in this paper has been supported in part by the NSFC project (61572376).

%\newpage
%\bibliographystyle{named}
%\bibliography{ijcai17_poi}

\end{document}